\documentclass[twoside,fleqn]{article}

\usepackage{epsf}
\usepackage{espcrc2}
\usepackage{amssymb}

\def\CO{{\ensuremath{\cal O}}}

\def\CW{{\ensuremath{\cal W}}}
\def\slash#1{\ensuremath{\mbox{$\not \!\! #1$}}}

\title{
       \begin{flushright}\normalsize
	    \vskip -0.9 cm
            LA-UR-00-5683
       \end{flushright}
	\vskip -0.4 cm
	Improvement and Renormalization Constants in $O(a)$ Improved Lattice QCD \thanks{
	Presented by Rajan Gupta. Calculations supported by the DoE Grand Challenges award 
	at the ACL at Los Alamos and NERSC.}
        }
\author{Tanmoy Bhattacharya\address{MS B-285, 
		Los Alamos National Lab, Los Alamos, New Mexico 87545, USA}, 
        Rajan Gupta$\null^{\rm a}$, 
	Weonjong Lee$\null^{\rm a}$,
	Stephen Sharpe\address{Physics Department, University of Washington,
         Seattle, Washington 98195, USA}
       }

\begin{document}

\begin{abstract}
We present results at $\beta=6.0$ and $6.2$ for the $O(a)$ improvement
and renormalization constants for bilinear operators using axial and
vector Ward identities. We discuss the extraction of the mass
dependence of the renormalization constants and the coefficients of 
the equation of motion operators.
\end{abstract}

\maketitle

\section{INTRODUCTION}

In quenched Lattice QCD, axial and vector Ward identities can be used
to determine, at $O(a)$, all the scale independent renormalization
constants for bilinear currents ($Z_A$, $Z_V$, and $Z_S/Z_P$), the
improvement constants ($c_A$, $c_V$, and $c_T$), the quark mass
dependence of all five $Z_{\cal O}$, and the coefficients of the
equation of motion operators
~\cite{ALPHA:Zfac:96,ALPHA:Zfac:97A,ALPHA:Zfac:97B,Bochicchio:cs:85,LANL:Zfac:98,LANL:Zfac:00}.  
Here we summarize
results at $\beta=6.0$ and $6.2$ and discuss the highlights of our
calculations. Details and notation are given in ~\cite{LANL:Zfac:00}.\looseness-1

\begin{table*}
\begin{center}
\caption{Simulation parameters, statistics, and the time
interval in $x_4$ defining the volume $V$ over which the chiral rotation is
performed in the AWI\@. The lattice spacing is fixed using $r_0=0.5$ fermi, and 
is thus independent of the fermion action. 
The source $J$ is placed at $t=0$.}
\begin{tabular}{|l|c|c|c|c|c|c|c|}
\hline
\multicolumn{1}{|c|}{Label}&
\multicolumn{1} {c|}{$\beta$}&
\multicolumn{1} {c|}{$c_{SW}$}&
\multicolumn{1} {c|}{$a^{-1}$ (GeV)}&
\multicolumn{1} {c|}{Volume}  &
\multicolumn{1} {c|}{$L$ (fm)}&
\multicolumn{1} {c|}{Confs.}  &
\multicolumn{1} {c|}{$x_4$}  \\
\hline
{\bf 60TI}   & 6.0 & 1.4755 & 2.12      & $16^3 \times 48$  & 1.5  & 83       & $4 - 18$  \\
\hline       	     
{\bf 60NPf}  & 6.0 & 1.769  & 2.12      & $16^3 \times 48$  & 1.5  & 125      & $4 - 18$  \\
{\bf 60NPb}  &     &        &           &                   &      & 112      & $27 - 44$ \\
\hline       	     
{\setbox0=\vbox{\hrule width 0pt\relax
 \vskip 5pt\hbox{\bf 62NP }}\dp0=0pt\relax
 \ht0=0pt\relax\box0}
& 6.2 & 1.614  & 2.91      & $24^3 \times 64$  & 1.65 & 70       & $6 - 25$  \\
             &     &        &           &                   &      & 70       & $39 - 58$ \\
\hline    
\end{tabular}
\end{center}
\label{tab:lattices}
\vskip -0.4 cm
\end{table*}

We start with the generalized axial Ward identity involving operators
improved on and off-shell
%
\[
\langle \delta S^{(12)} \ \CO^{(23)}_{R,\it off} (y) \ J^{(31)}(0) \rangle 
= \langle \delta \CO^{(13)}_{R,\it off} (y) \ J^{(31)} (0) \rangle 
\label{eq:AWI1}
\]
%
%
where $\delta\CO$ is the result of the axial variation of $\CO$
($A_\mu \leftrightarrow V_\mu$, $S \leftrightarrow P$, and $T_{\mu\nu}
\rightarrow \epsilon_{\mu\nu\rho\sigma}T_{\rho\sigma}$), and $\delta S$ is 
the variation in the action. 

At $O(a)$ there exists only one dimension 4 off-shell operator
(which vanishes by the equations of motion) for each bilinear that has the
appropriate symmetries~\cite{ROME:Imp:97}. Consequently, we define
\begin{eqnarray}
\CO_{R,\it off}^{(ij)} &=&   Z_\CO^{(ij)}  \CO_{I,\it off}^{(ij)} \,,
\\
\CO_{I,\it off}^{(ij)} &=& \CO_I^{(ij)} - (1/4)\ a  c'_\CO E_\CO^{(ij)} \,,
\\
E_\CO^{(ij)} &=& \bar{\psi}^{(i)}  \Gamma \overrightarrow{\CW}  \psi^{(j)} - 
		 \bar{\psi}^{(i)}  \overleftarrow{\CW}  \Gamma \psi^{(j)} \,.
\label{eq:EMdef}
\end{eqnarray}
Here $(ij)$ (with $i\ne j$) specifies the flavor, and 
$\overrightarrow{ \CW} \psi_j =
(\overrightarrow{\slash{D}}+m_j)\psi_j +O(a^2)$ is the full $O(a)$
improved Dirac operator for quark flavor $j$. This ensures that the
equation-of-motion operator $E_\CO$ gives rise only to contact terms,
and does not change the overall normalization $Z_\CO$. The $O(a)$ 
on-shell improved renormalized operators $\CO_R^{(ij)}$ are 
\begin{eqnarray}
\CO_R^{(ij)}    & \equiv & Z_\CO^0(1+b_\CO am_{ij} ) \CO_I^{(ij)} \,, \\
                & \equiv & Z_\CO^0(1+{\tilde b}_\CO a {\tilde m}_{ij} ) \CO_I^{(ij)} \,.
\label{eq:Ordef} 
\\
(A_{I})_{\mu}    & \equiv & A_{\mu} + a c_A \partial_\mu P  \,,  \\
(V_{I})_{\mu}    & \equiv & V_{\mu} + a c_V \partial_\nu T_{\mu\nu} \,,  \\
(T_{I})_{\mu\nu} & \equiv & T_{\mu\nu} +
                a c_T ( \partial_\mu V_\nu - \partial_\nu V_\mu) \,,  \\
P_I &\equiv& P \,, \qquad S_I \equiv S \,,
\label{eq:impbilinears}
\end{eqnarray}
The $Z_\CO^0$ are renormalization constants in the chiral limit,
$m_{ij} \equiv ( m_i + m_j)/2$ is the average bare quark mass, $a
m_i=1/2\kappa_i - 1/2\kappa_c$, $\kappa_c$ is the value of the hopping
parameter in the chiral limit, and ${\tilde m}_{ij}$ is the quark mass
defined by the axial Ward identity (AWI) in Eq.~\ref{cA}. Note that $m$ and
$\tilde m$ are identical in a discretized theory with chiral symmetry,
like staggered fermions.  With these definitions, $b_\CO=1$,
$c_\CO=0$, $c'_\CO=1$ at tree level~\cite{LANL:Zfac:00}.

Since the equation-of-motion operators contribute only contact terms, 
Eq.~(\ref{eq:AWI1}) can be rewritten in terms of just on-shell improved 
operators:
\begin{eqnarray}
& & \frac{
	\langle \int_V d^4x\, \delta S_I
	\ \CO_I^{(23)}(y_4, \vec y) \ J^{(31)}(0)
	\rangle }
	{  \langle 
	\delta \CO_I^{(13)}(y_4, \vec y) \ J^{(31)}(0)
	\rangle }  \nonumber \\
&=&  \frac{ Z^{(13)}_{\delta\CO} }
	{ Z^{(12)}_A \ Z^{(23)}_\CO }
	+ a \frac{c'_P + c'_\CO}{2} {\tilde m}_{12} + O(a^2)
\label{eq:WI-c'}
\end{eqnarray}
where
\[
\delta S_I (x) \equiv
2{\tilde m}_{12} P_I^{(12)}(x)
	- \partial_\mu (A_I)^{(12)}_\mu(x)  \,.
\label{eq:deltaSI}
\]
Our calculation is limited to the case ${\tilde m}_1 = {\tilde m}_2$ 
(this simplification was used due to limited computer resources), 
in which case the r.h.s. of Eq.~\ref{eq:WI-c'} reduces to 
\begin{eqnarray}
& & {\hskip -0.7 cm } 
        \frac{ Z^0_{\delta\CO} } { Z^0_A \ Z^0_\CO } 
	\Big[ 1 +  (\tilde b_{\delta \CO} - \tilde b_\CO ) 
	\frac{a \tilde m_3}{2} \Big] +  
	\nonumber \\ 
& & {\hskip -0.7 cm } 
	 \bigg[ \frac{ Z^0_{\delta\CO} } { Z^0_A \ Z^0_\CO } 
         \Big( \frac{ ( \tilde b_{\delta \CO} - \tilde b_\CO ) }{2} 
		- \tilde b_A \Big)
	+ \frac{c'_P + c'_\CO}{2} \bigg] a \tilde m_1 
\label{eq:AWIrhs2}
\end{eqnarray}
where $\tilde m_i \equiv {\tilde m}_{ij}|_{m_j=m_i}$.  Using  Eqs.~\ref{eq:WI-c'} and 
\ref{eq:AWIrhs2}, all
the $b_\CO$ (except $b_T$ which requires $m_1 \neq m_2$), $c_\CO$,
$c'_\CO$, and the scale independent normalization constants are
determined by making suitable choices for $J$, $\CO$, and $y$ in
Eq.~\ref{eq:WI-c'} and studying it as a function of ${\tilde m}_1$ and
${\tilde m}_3$ (Eq.~\ref{eq:AWIrhs2}).

\section{RESULTS}

The lattice parameters used in our calculation are given in
Tab.~\ref{tab:lattices} and \ref{tab:qmasses}.  Our final results,
which supercede those in Ref.~\cite{LANL:Zfac:98}, are given in
Tab.~\ref{tab:finalcomp}. 

In many cases a given on-shell improvement
and normalization constants can be determined in a number of ways as
discussed in~\cite{LANL:Zfac:00}. Results in Tab.~\ref{tab:finalcomp}
are based on the AWI with the best signal and smallest error.
Table~\ref{tab:finalcomp} also includes results by the ALPHA
collaboration~\cite{ALPHA:Zfac:96,ALPHA:Zfac:97A,ALPHA:Zfac:97B} and
the one-loop tadpole improved perturbative results.  To simplify
comparison with previous results, we quote both \({\tilde b}_V,
{\tilde b}_A\) and \(b_V, b_A\).

\begin{table*}
\begin{center}
\caption{Values of $\kappa$ used in the three simulations, and the
corresponding values of $a M_\pi$ and the quark mass $a \tilde m$
extracted.  $\tilde m$ is defined by the AWI in
Eq.~\protect\ref{cA}. $\kappa_c$ is the zero of $\tilde m$ obtained from
quadratic fits in $1/\kappa$. The non-zero value of $a M_\pi$ at $\kappa_c$ is 
indicative of the inadequacy of quadratic fits, $a^2 M_\pi^2 $ as a function of $1/2\kappa$, 
used to extract it, and discretization errors. Of these, the first is the dominant cause and 
points to the need for including quenched chiral logs in the fits~\protect\cite{LANL:Zfac:00}. }
\advance\tabcolsep by -3.3pt
\begin{tabular}{|c|c|c|c|c|c|c|c|c|c|}
\hline
\multicolumn{1}{|c|}{}&
\multicolumn{3} {c|}{{\bf 60TI}}&
\multicolumn{3} {c|}{{\bf 60NP}}&
\multicolumn{3} {c|}{{\bf 62NP}}\\
\multicolumn{1}{|c|}{Label}  &
\multicolumn{1} {c|}{$\kappa$}  &
\multicolumn{1} {c|}{$a{\tilde m}$}  &
\multicolumn{1} {c|}{$aM_\pi $}  &
\multicolumn{1} {c|}{$\kappa$}  &
\multicolumn{1} {c|}{$a{\tilde m}$}  &
\multicolumn{1} {c|}{$aM_\pi$}  &
\multicolumn{1} {c|}{$\kappa$}  &
\multicolumn{1} {c|}{$a{\tilde m}$}  &
\multicolumn{1} {c|}{$a M_\pi $}  \\
\hline    
%
%
$\kappa_1$ & $0.11900$ & $0.443(8)$ & $1.530(1)$ & $0.1300$ & $0.144(1)$ & $0.711( 2)$ &  $0.1310$ & $0.1345(6)$  & $0.609(1)$ \\
$\kappa_2$ & $0.13524$ & $0.105(1)$ & $0.571(2)$ & $0.1310$ & $0.118(1)$ & $0.630( 2)$ &  $0.1321$ & $0.1054(4)$  & $0.522(1)$ \\
$\kappa_3$ & $0.13606$ & $0.084(1)$ & $0.504(2)$ & $0.1320$ & $0.092(1)$ & $0.544( 2)$ &  $0.1333$ & $0.0727(3)$  & $0.418(1)$ \\
$\kappa_4$ & $0.13688$ & $0.063(1)$ & $0.431(2)$ & $0.1326$ & $0.075(1)$ & $0.488( 2)$ &  $0.1339$ & $0.0560(2)$  & $0.360(2)$ \\
$\kappa_5$ & $0.13770$ & $0.042(1)$ & $0.348(3)$ & $0.1333$ & $0.056(1)$ & $0.416( 2)$ &  $0.1344$ & $0.0419(2)$  & $0.307(2)$ \\
$\kappa_6$ & $0.13851$ & $0.020(1)$ & $0.244(4)$ & $0.1342$ & $0.032(1)$ & $0.308( 3)$ &  $0.1348$ & $0.0306(2)$  & $0.261(2)$ \\
$\kappa_7$ & $0.13878$ & $0.013(1)$ & $0.195(8)$ & $0.1345$ & $0.025(4)$ & $0.262(12)$ &  $0.1350$ & $0.0248(1)$  & $0.235(2)$ \\
$\kappa_c $      &$0.13926(2)$  &  0      & $0.082(15)$ & 
                  $0.13532(3)$  &  0      & $0.083(20)$ & 
                  $0.135861(5)$ &  0      & $0.066(10)$ \\
\hline    
\end{tabular}
\end{center}
\label{tab:qmasses}
\vskip -0.4 cm
\end{table*}

One of the goals of our calculation is to quantify the residual
$O(a^2)$ errors and to understand the shortcomings of 1-loop
perturbation theory. For $O(a^2)$ errors we use two estimates: (i) the
difference between our results and those by the ALPHA
collaboration~\cite{ALPHA:Zfac:96,ALPHA:Zfac:97A,ALPHA:Zfac:97B}, and
(ii) the difference between using 2-point and 3-point discretization
of the derivatives~\cite{LANL:Zfac:00} in the extraction of $c_A$ from
\begin{eqnarray}
& &    \frac{ \sum_{\vec{x}} \langle 
       \partial_\mu [A_\mu + 
       a c_A \partial_\mu P]^{(ij)}(\vec{x},t) J^{(ji)}(0) \rangle} 
       {\sum_{\vec{x}} \langle P^{(ij)}(\vec{x},t) J^{(ji)}(0) \rangle} \nonumber \\
 &=& 2 {\tilde m}_{ij}   \,,
\label{cA}
\end{eqnarray}
and the subsequent effect of the difference in $c_A$ on other constants.  This
latter variation is quoted as the second error in Tab.~\ref{tab:finalcomp}.

These differences are compared to the expected size of the residual 
discretization errors: $(a\Lambda_{\rm QCD})\approx0.15$ and $0.1$ for
the improvement constants and $(a\Lambda_{\rm QCD})^2\approx0.02$ and
$0.01$ for the normalization constants at $\beta=6.0$ and $6.2$
respectively.

A comparison, at $\beta=6.0$, between simulation at $c_{SW}=1.4755$
(tadpole improved theory) and $c_{SW}=1.769$ (non-perturbatively
$O(a)$ improved theory) shows that all the constants are sensitive to
the choice of $c_{SW}$. It is therefore important to use $c_{SW}$
determined non-perturbatively.

The most significant comparison is between our results and those of
the ALPHA collaboration.  The only results which do not agree within
2-$\sigma$ statistical errors are those for $Z_V^0$, $c_A$ and $c_V$
at $\beta=6$, and for $Z_V^0$ at $\beta=6.2$. The differences for
$Z_V^0$ are of size $0.01$ and $0.005$ at $\beta=6$ and $6.2$
respectively, and are thus consistent with the expected differences of
$O(a^2)$.  The differences for $c_A$ and $c_V$ are also consistent
with the size expected of $O(a)$ differences, but are more notable
because they correspond to very large fractional differences (e.g. our
$c_A$ at $\beta=6$ has less than half the magnitude of that found by
the ALPHA collaboration).  What we learn is that (i) $c_\CO$, which
vanish at tree level and are numerically small, depend
substantially, at $\beta=6$, on the method/definition used to extract
them; (ii) the variation between 2-pt and 3-pt derivatives
significantly smaller than the difference between our results and
those of the ALPHA collaboration; and (iii) these differences in
$c_V$, and even more so in $c_A$, are substantially reduced at
$\beta=6.2$.  The change appears too rapid to be an $O(a)$ effect.

Both $c_V$ and $c_T$ are obtained as a small difference between two
large terms.  Nevertheless, we are able to design Ward identities that
yield these quantities with reasonable precision. In particular, the
significant improvement we obtain in determining $c_V$ using methods
described in \cite{LANL:Zfac:00} reduces the error in $Z_A^0$,
$Z_P^0/Z_S^0$, $c_T$ and $c'_A$ as the uncertainty in $c_V$ feeds into
these quantities.

\begin{table*}[ht]
\begin{center}
\caption{Final results for improvement and renormalization constants.
The first error is statistical, and the second, where present,
corresponds to the difference between using 2-point and 3-point
discretization of the derivative used in the extraction of $c_A$. The
ALPHA collaboration results are
from~\cite{ALPHA:Zfac:96,ALPHA:Zfac:97A,ALPHA:Zfac:97B}.  For the
tadpole improved perturbative results (labelled P.Th.) see appendix in
Ref.~\cite{LANL:Zfac:00} and references there.  }
\setlength{\tabcolsep}{3.7pt}
\begin{tabular}{||c||l|l|l|l||l|l|l||}
\hline
\multicolumn{1}{||c||}{}&
\multicolumn{4}{c||}{\(\beta=6.0\)}&
\multicolumn{3}{c||}{\(\beta=6.2\)}\\
\hline
         & LANL              & LANL             & ALPHA        & P. Th.
         &  LANL             &  ALPHA           &  P. Th.     \\
         &                   &                  &              &
         &                   &                  &                \\[-12pt]
\hline			     		       
         &                   &                  &              &
         &                   &                  &                \\[-12pt]
$c_{SW}$ & 1.4755            & 1.769            & 1.769        &  1.521
         &  1.614            &  1.614           &  1.481         \\
         &                   &                  &              &
         &                   &                  &                \\[-12pt]
\hline			     		       
         &                   &                  &              &
         &                   &                  &                \\[-12pt]
$Z^0_V$  & $+0.747(1)   $    & $+0.770(1)    $  & $0.7809(6)$  & $+0.810$  
         & $+0.7874(4)  $    & $+0.7922(4)(9)$  &  $+0.821$   \\
$Z^0_A$  & $+0.791(7)(4)$    & $+0.807(2)(8) $  & $0.7906(94)$ & $+0.829$  
         & $+0.818(2)(5)$    & $+0.807(8)(2) $  &  $+0.839$   \\
$Z^0_P/Z^0_S$		     		       				    
         & $+0.811(9)(5)$    & $+0.842(5)(1)$    &  N.A.        & $+0.956$  
         & $+0.884(3)(1)$    &  N.A.            & $+0.959$    \\
         &                   &                  &              &              
         &                   &                  &                \\[-12pt]
\hline			     		       
         &                   &                  &              &              
         &                   &                  &                \\[-12pt]
$c_A$    & $-0.022(6)(1)$    & $-0.037(4)(8)$   & $-0.083(5)$  & $-0.013$  
         & $-0.032(3)(6)$    &  $-0.038(4)$     &  $-0.012$   \\
$c_V$    & $-0.25 (5)(3)$    & $-0.107(17)(4)$  & $-0.32 (7)$  & $-0.028$  
         & $-0.09 (2)(1)$    &  $-0.21(7)$      &  $-0.026$   \\
$c_T$    & $+0.09 (2)(1)$    & $+0.06 (1)(3)$   &  N.A.        & $+0.020$  
         & $+0.051(7)(17)$   &  N.A.            &  $+0.019$   \\
         &                   &                  &              &              
         &                   &                  &                \\[-12pt]
\hline			     		       
         &                   &                  &              &              
         &                   &                  &                \\[-12pt]
$\tilde b_V$		     		       				    
         & $+1.44 (3)(2)$    & $+1.43(1)(4)$    &  N.A.        & $+1.106$ 
         & $+1.30 (1)(1)$    &  N.A.            &  $+1.099$  \\
$b_V$    & $+1.53 (2)$       & $+1.52(1)$       & $+1.54(2)$   & $+1.274$ 
         & $+1.42 (1)$       & $+1.41(2)$       &  $+1.255$  \\
$\tilde b_A-\tilde b_V$	     		       				    
         & $-0.51 (9)(4)$    & $-0.26(3)(4)$    &  N.A.        & $-0.002$  
         & $-0.11 (3)(4)$    &  N.A.            &  $-0.002$   \\
$b_A-b_V$	     		       				    
         & $-0.49 (9)(4)$    & $-0.24(3)(4)$    &  N.A.        & $-0.002$  
         & $-0.11 (3)(4)$    &  N.A.            &  $-0.002$   \\
$\tilde b_P-\tilde b_S$	     		       				    
         & $-0.07 (9)(2)$    & $-0.06(4)(3)$    &  N.A.        & $-0.066$  
         & $-0.09 (2)(1)$    &  N.A.            &  $-0.062$   \\
$\tilde b_P-\tilde b_A$	     		       				    
         & $-0.126(58)(1)$   & $-0.07(4)(5)$    &  N.A.        & $+0.002$  
         & $-0.09 (3)(3)$    &  N.A.            &  $+0.001$   \\
         &                   &                  &              &              
         &                   &                  &                \\[-12pt]
\hline			     		       
         &                   &                  &              &              
         &                   &                  &                \\[-12pt]
$\tilde b_A$		     		       				    
         & $+0.92 (10)(6)$   & $+1.17(4)(8)$    &  N.A.        & $+1.104$ 
         & $+1.19 (3)(5)$    &  N.A.            &  $+1.097$   \\
$b_A$		     		       				    
         & $+1.05 (9)(4)$    & $+1.28(3)(4)$    &  N.A.        & $+1.271$ 
         & $+1.32 (3)(4)$    &  N.A.            &  $+1.252$   \\
$\tilde b_P$		     		       				    
         & $+0.80 (11)(6)$   & $+1.10(5)(13)$   &  N.A.        & $+1.105$ 
         & $+1.11 (4)(7)$    &  N.A.            &  $+1.099$  \\
$\tilde b_S$		     		       				    
         & $+0.87 (14)(4)$   & $+1.16(6)(11)$  &  N.A.        & $+1.172$ 
         & $+1.19 (4)(6)$    &  N.A.            &  $+1.161$   \\[3pt]
\hline
\end{tabular}
\label{tab:finalcomp}
\end{center}
\vskip -0.3cm
\end{table*}

When comparing against perturbative estimates the yardstick we use for
the missing higher order terms is $\sim \alpha_s^2\approx 0.02$ and
$0.016$, respectively.  We find that tadpole-improved
1-loop perturbation theory underestimates the deviations of
renormalization and improvement constants from their tree level
values.  In all but one case, however, these discrepancies can be
understood as a combination of a 2-loop correction of size $(1-2)
\times \alpha_s^2$ [for $Z_V^0$, $Z_A^0$, and $c_A$], higher order
discretization errors of size $(1-2)\times a\Lambda_{\rm QCD}$ [for
$c_V$, $c_T$ and ${\tilde b}_V$], and statistical errors [for ${\tilde
b}_A$, ${\tilde b}_P$, and ${\tilde b}_S$].  The only exception is
$Z_P^0/Z_S^0$, for which a very large higher order perturbative
contribution of size $4\times\alpha_s^2$ is needed to reconcile our
non-perturbative results with 1-loop perturbation theory.

In Tab.~\ref{tab:c'}, we present, first results for the
equation-of-motion improvement constants $c_X^\prime$.  The
combination ${c'_P + c'_\CO}$ is extracted by studying the dependence
of Eq.~\ref{eq:WI-c'} on ${\tilde m}_1$ once the other constants
defined in Eq.~\ref{eq:AWIrhs2} have been determined. The errors in
the determination of the $c'_\CO$ are dominated by two quantities: (i)
The uncertainty in $c_A$ feeds into the extraction of $c'_A$, and (ii)
the correlation function from which ${c'_P + c'_P}$ is determined has
a poor signal (the intermediate state is a scalar for $ J=S $, $ \CO =
P $ and $ \delta \CO = S $ in Eq.~\ref{eq:WI-c'}). The uncertainty in
$c'_P$ then feeds into $c'_V$, $c'_S$, and $c'_T$.  Overall, we find a
very significant improvement in the quality of the results with
increasing $\beta$, $i.e.$, between $\beta=6.0$ and $\beta=6.2$.

\begin{table}
\caption{Results for off-shell mixing coefficients.}
\setlength{\tabcolsep}{1.5pt}
\renewcommand{\arraystretch}{1.2}
\begin{center}
\begin{tabular}{|c|c|c|c|c|}
\hline
\multicolumn{1}{|c|}{}&
\multicolumn{1} {c|}{\bf 60TI}&
\multicolumn{1} {c|}{\bf 60NPf}&
\multicolumn{1} {c|}{\bf 60NPb}&
\multicolumn{1} {c|}{\bf 62NP}\\
\hline
 $c'_V     $  &  $ +3.72 (  73 )  $ &  $ +2.38 (  50 )  $ &  $ +3.00 (  37 )  $ &  $ +1.72 (  16 )  $  \\ 
 $c'_A     $  &  $ +3.28 (  94 )  $ &  $ +1.99 (  56 )  $ &  $ +2.45 (  46 )  $ &  $ +1.53 (  20 )  $  \\ 
 $c'_P     $  &  $ -0.98 (  76 )  $ &  $ +0.44 (  49 )  $ &  $ -0.33 (  29 )  $ &  $ +0.91 (  12 )  $  \\ 
 $c'_S     $  &  $ +3.00 (  73 )  $ &  $ +2.00 (  48 )  $ &  $ +2.72 (  33 )  $ &  $ +1.49 (  14 )  $  \\ 
 $c'_T     $  &  $ +3.24 (  75 )  $ &  $ +1.96 (  49 )  $ &  $ +2.60 (  38 )  $ &  $ +1.51 (  15 )  $  \\   
\hline
\end{tabular}
\label{tab:c'}
\end{center}
\vskip -0.8cm
\end{table}

Finally, we comment on results presented in two recent papers. Using
the Schr\"odinger functional, Ref.~\cite{guagnelli2000} calculates
$b_A-b_P$, $b_S$ and $Z_P^0/(Z_S^0 Z_A^0)$ for a range of $\beta \ge
6$.  Their most striking result is that different discretizations of
derivatives lead to very different results for $b_A-b_P$.  For
example, at $\beta=6$, this quantity varies roughly from $0.17$ to
$-0.17$. While our number lies within this range, our estimate of
$O(a)$ uncertainties is clearly a substantial underestimate.  

Reference~\cite{collins2000} has determined $c_A$ using the same
method and similar lattice parameters as here but with significantly
more configurations.  They study, at one $\kappa$ ($\sim \kappa_5$ at
both $\beta=6.0$ and $6.2$), the effect of using derivatives that are
tree-level improved through $O(a^2)$ (our 3-pt), $O(a^4)$ and
$O(a^6)$. They find a larger dependence than what we get between 2-pt
and 3-pt discretizations at $\kappa_5$. The $O(a^2)$ errors in the two
calculations are, however, different due to the choice of source and
the fit range in time. Also, we find that after chiral extrapolation
these discretization effects are significantly reduced. Nevertheless,
once again the large variation should serve as a warning that the
$O(a)$ errors in $c_\CO$ can be substantial.


%

\def\ie{{\sl i.e.}}
\def\etal{{\it et al.}}
\def\etc{{\it etc.}}
\def\ibid{{\it ibid}}


\end{document}